\documentclass[9.5pt,journal,compsoc]{IEEEtran}

\usepackage{tcolorbox}
\usepackage[utf8]{inputenc}
\usepackage{array,multirow,etoolbox,graphicx,subcaption,fancyhdr,tabularx,longtable}
\usepackage{float}
\usepackage{graphicx}
\usepackage{amsmath} 
\usepackage{tcolorbox}
\usepackage{booktabs}
\usepackage{mathtools}
\usepackage{xcolor}
\usepackage{hyperref}
\newtcolorbox{TcolorBox}[1]{fonttitle=\bfseries,title=#1}
\usepackage{csquotes}
\usepackage[euler]{textgreek}
\usepackage{microtype}
\usepackage{tablefootnote}
\usepackage{threeparttable}
\usepackage{listings}
\newcommand{\crres}{{\tt MetaMateCR}\xspace}
\newcommand{\actionable}{ActionableToApplied\xspace}
\newcommand{\shown}{ShownToApplied\xspace}

\newcommand{\SftSmall}{{\tt SmallLSFT}\xspace}
\newcommand{\SftLarge}{{\tt LargeLSFT}\xspace}
\newcommand{\timeInReview}{TimeInReview\xspace}
\newcommand{\eyeballTime}{TimeSpent\xspace}
\newcommand{\wallClock}{WallClock\xspace}

\newlength{\imagewidth}
\newcommand{\imagescale}{.30}

\usepackage{amsmath,amsfonts}
\usepackage{algorithmic}
\usepackage{graphicx}
\usepackage{textcomp}
\usepackage{xcolor}
\usepackage{url}
\usepackage{microtype}

\usepackage{xurl} 
\usepackage{tabularx}
\usepackage{caption}
\usepackage{subcaption}
\usepackage{xspace}

\usepackage{enumitem}
\newlist{steps}{enumerate}{1}
\setlist[steps, 1]{label = Step \arabic*:}

\usepackage{xspace}
\usepackage{dirtytalk}
\usepackage{xcolor}

\newcommand{\company}{Meta\xspace}
\newcommand{\Meta}{Meta\xspace}

\usepackage{xspace}
\usepackage{dirtytalk}
\usepackage{xcolor}

\definecolor{teal}{RGB}{0,128,128}

\newcommand{\etal}{\hbox{\emph{et al.}}\xspace}
\newcommand{\eg}{\hbox{\emph{e.g.,}}\xspace}
\newcommand{\ie}{\hbox{\emph{i.e.}}\xspace}

\newcommand{\DefMacro}[2]{\expandafter\newcommand\csname rmk-#1\endcsname{#2}}
\newcommand{\UseMacro}[1]{\csname rmk-#1\endcsname}
\newcommand{\rom}[1]{\uppercase\expandafter{\romannumeral #1\relax}}

\newcommand{\gpt}{{\tt GPT-4o}\xspace}

\begin{document}
\title{AI-Assisted Fixes to Code Review Comments at Scale}

\author{Chandra Maddila, Negar Ghorbani, James Saindon, Parth Thakkar, Vijayaraghavan Murali, Rui Abreu, Jingyue Shen, Brian Zhou, Nachiappan Nagappan, and Peter~C. Rigby~\IEEEmembership{}
\IEEEcompsocitemizethanks{\IEEEcompsocthanksitem All the authors are with Meta Platform Inc, Menlo Park, California, USA. P. C. Rigby is also with the Department of Computer Science and Software Engineering, Concordia University, Montr{\'e}al, Qu{\'e}bec, Canada.\protect\\
Corresponding Author E-mails: cmaddila@meta.com, pcr@meta.com}
}

\markboth{Transactions on Software Engineering,~Vol.~X, No.~Y, July~2025}%
{Shell \MakeLowercase{\textit{Maddila \etal}}: AI-Assisted Fixes to Code Review Comments at Scale}

\IEEEtitleabstractindextext{
{\bf Aim.} There are $10$s of thousands of code review comments each week at \Meta. We developed Metamate for Code Review (\crres) that provides AI-assisted fixes for reviewer comments in production at scale.

{\bf Method.} We developed an internal benchmark of $64$k $\langle review\_comment, patch \rangle$ data points to fine-tune Llama models. Once our models achieve reasonable offline results, we roll them into production. To ensure that our AI-assisted fixes do not negatively impact the time it takes to do code reviews, we conduct randomized controlled safety trials as well as full production experiments.

{\bf Offline Results.} As a baseline, we compare \gpt to our small and large Llama models. In offline results, our \SftLarge model creates an exact match patch 68\% of the time outperforming \gpt by 9 percentage points (pp). The internal models also use more modern Hack functions when compared to the PHP functions suggested by \gpt.

{\bf Safety Trial.} When we roll \crres into production in a safety trial that compares no AI patches with AI patch suggestions, we see a large regression with reviewers taking over 5\% longer to conduct reviews. After investigation, we modify the UX to only show authors the AI patches, and see no regressions in the time for reviews. 

{\bf Production.} When we roll \SftLarge into production, we see an \actionable rate of 19.7\%, which is a 9.2pp improvement over \gpt. 

Our results illustrate the importance of safety trials in ensuring that AI does not inadvertently slow down engineers, and a successful review comment to AI patch product running at scale.

\begin{IEEEkeywords}
       Code review, AI patches, Supervised Fine Tuning, Randomized Controlled Safety Trials, Production Experiments
\end{IEEEkeywords}
}

\IEEEdisplaynontitleabstractindextext

\IEEEpeerreviewmaketitle

\maketitle

\section{Introduction}
Code review is important to ensure that high quality code is released into production. The process has evolved from a heavyweight inspection process to increasingly lightweight tool-based approaches that rely on reviewer expertise~\cite{Fagan1976IBM,porter1998understanding,rigby2013convergent}.
With the rise of generative models (LLMs), code review research has shifted towards automating more complex activities. For example, Cihan \etal~\cite{cihan2024automated} propose a method to do LLM-based code review and evaluate it in an enterprise setting. AutoTransform~\cite{autotransform2024} proposed a neural-network based method for transforming code during the code review process.

Recent works have used LLMs to generate code fixes from reviewer comments. For example, Google~\cite{google2024cr2} reports on the creation of a production system. Over the last few years, we have also been developing and deploying Metamate for Code Review (\crres). 
We discuss the motivation for each of our models, using an illustrative example of a reviewer's comment, and the AI patch fixes in Figure~\ref{figIllustrationModels}. The figure also shows the UX of Metamate for Code Review (\crres) in production. We conduct safety trials to ensure that AI fix suggestions do not inadvertently slow down code review, and also rollout at scale. 

Below we summarize our motivation and briefly highlight our contributions through our research results.

\subsection{\gpt: Public model}
When we started this research (in 2024), \gpt was the state-of-the-art public model for coding.
We conducted an offline test on an internal benchmark using \gpt. We calculate how often the model produces the exact match (EM) patch from the comment and the number times that it creates a patch that can compile, that does not contain static analysis and lint warnings, and passes all tests (SPG). 

\noindent \textbf{Results summary.} 
As illustrated in Figure~\ref{figIllustrationModels}a, we see that \gpt is able to correctly fix the patch, but because it only has publicly available information, it uses an outdated PHP function rather than the modern performant Hack function.  
Our offline evaluation shows that \gpt generates an exact match 59\% of the time. When we roll it out into production, we find that authors accept the patch 13\% of the time. 

\subsection{\SftSmall: Llama-SFT on on a small human-labeled data set}

To fine-tune a model for our internal context, we turn to the Llama-3-70B. This open source public model is trained on both public documents and source code and serves as our foundation model for subsequent work \cite{grattafiori2024llama3herdmodels}.
Using this model, we pre-train on \company's internal code to create iCodeLlama-3-70B. We then curate a small high-quality human-labeled annotation data set and use supervised fine tuning (SFT) on iCodeLlama-3-70B. We call this model \SftSmall.

\noindent \textbf{Results summary.}  
As illustrated in Figure~\ref{figIllustrationModels}b, we see that \SftSmall uses a performant Hack function because it has been fine-tuned on the internal codebase. 
We find that \SftSmall has an offline exact match of 63\% and, in production experiments, authors accept the patch 16.13\%. In percentage points (pp), this represents a 4 pp and 1.25 pp improvement over \gpt. 

\subsection{\SftLarge: Llama-SFT on a large LLM-labeled code review data set}

It is difficult to manually curate a large training set of code review comments and appropriate change. 
We use the manually created \SftSmall data set to train iCodeLlama-3-8B as a classifier to curate a large SFT dataset. We are able to obtain 64k $\langle review\_comment, patch \rangle$ datapoints. We train iCodeLlama-3-70B on this large dataset to create \SftLarge.

\noindent \textbf{Results summary.} 
As illustrated in Figure~\ref{figIllustrationModels}c, we see that \SftLarge not only avoids PHP and uses performant Hack functions, but it writes clean code based on the large set of review comments that lead to accepted patches that it was trained on.
\SftLarge has an offline exact match of 68\% which is an improvement of 5pp and 9pp over \SftSmall and \gpt, respectively. The corresponding production value is the \actionable rate of 19.75\%, which is a 3.6 pp and 6.6 pp improvement over \SftSmall and \gpt, respectively.

\subsection{Safety Trials in Production}
Concurrently to the development of the models and offline tests, the engineering team designed a new user experience for \crres and conducted safety trials, similar to those used in medicine~\cite{jadad2007randomized}, to ensure that there were no unintended side effects of AI patch fixes. We conducted two randomized controlled safety experimental trials to ensure that AI patches from code review comments did not have a negative impact on the amount of time diffs spent in review, \timeInReview, and that reviewers did not spend more time actively examining diffs, \eyeballTime, and the total \wallClock from when the diff is published until it is closed. 

\noindent \textbf{Results summary.} The amount of time needed to do reviews increased by over 5\%. Investigation revealed that reviewers were checking the AI patches. In the second experiment, we only showed the AI suggestions to the authors and collapsed them for the reviewers. This change reduced the reviewer burden and led to a safe user experience that did not have any regressions in the time it takes for code review.

\subsection{Full Production Experiments}
Unlike medical trials~\cite{jadad2007randomized}, we are able to monitor the entire population of code reviews to ensure and continuously evaluate our goal and safety metrics~\cite{NudgeMeta}. We want all engineers to experience the best AI patch suggestion possible, so we rolled out the new models as they became available and monitored safety metrics as well as the goal metrics: \actionable and \shown

\noindent \textbf{Results summary.} In monitored production rollouts of \SftSmall and \SftLarge, we see substantial improvements over \gpt, 5.6pp and 9.22pp in \actionable, respectively. We monitor our safety metrics and do not experience any regressions.

The remainder of this paper is structured as follows. In Section~\ref{sec:background}, we provide background on \Meta, the code review process, and the automated validations processes. In Section~\ref{sec:modelDevelopment}, we introduce the foundation models and the fine-tuning that we use to develop our \crres models. In Section~\ref{secDataEvaluation}, we describe the process of developing our internal benchmark. Unfortunately, we cannot release the benchmark and data set, and, as a result, we do not consider the benchmark to be a research contribution. We also discuss our offline model evaluation, our randomized controlled trial, and our full production experimental methodologies. In Section~\ref{secOfflineResults}, we present the results from our offline model evaluation. In Section~\ref{secSystem}, we present the design and implementation of the production system. In Section~\ref{secOnlineResults}, we report the results for our online experiments and, ultimately, the rollout of \crres to engineers at \Meta. In Section~\ref{sec:threats}, we discuss the threats to validity of our model development and experiment. In Section~\ref{sec:literatureAndDiscussion}, we position and discuss our contribution in the research literature. In Section~\ref{sec:conclusion}, we conclude the work and emphasize our contributions and future directions.

\begin{figure*}
\centering
\caption{This figure illustrates the strengths and weaknesses of each of the patches generated from a code review comment by \gpt, \SftSmall, and \SftLarge. The human code reviewer comment (on the right with the username anonymized) suggests a change to check that the key (\texttt{encoder\char`_params}) exists in a dictionary (\texttt{vision\char`_encoder}) and then set it to a value if it is not present in the dictionary. In each subfigure, we discuss the patch fix suggested by each model.}

\subfloat[\gpt's response is functionally correct given the reviewer's comment. However, it is using an old PHP function (\texttt{encoder\char`_key\char`_exists}) to check if the key exists in a dictionary. This outdated PHP function is not as performant modern Hack functions.]{\label{g4-response}\includegraphics[width=.95\linewidth]{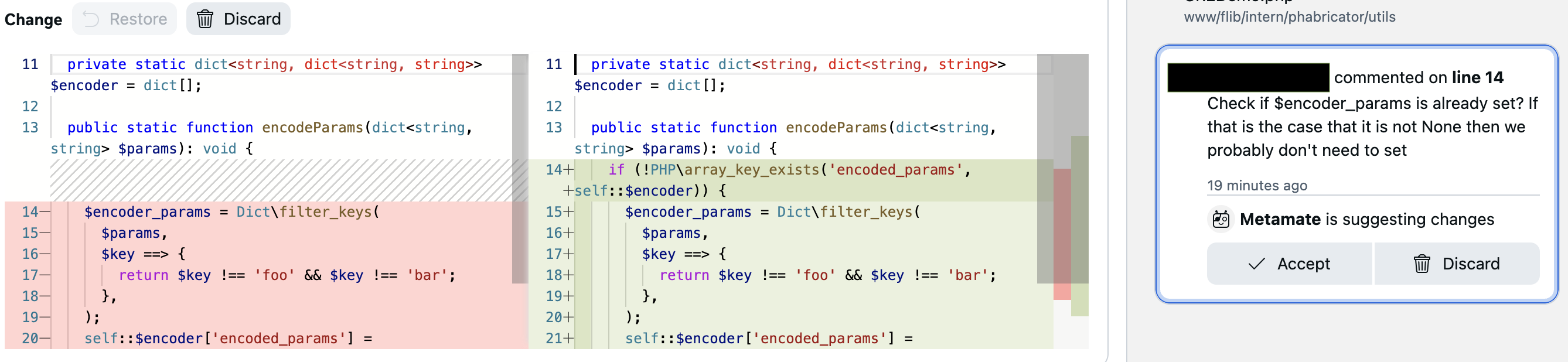}}\hfill

\subfloat[\SftSmall is finetuned on \Meta's internal codebase and reviewer comments. Its response to the reviewer's comment is to use the Hack function \texttt{C\char`\\contains\char`_key} and avoids outdated PHP functions.]{\label{smallSFT-response}\includegraphics[width=.95\linewidth]{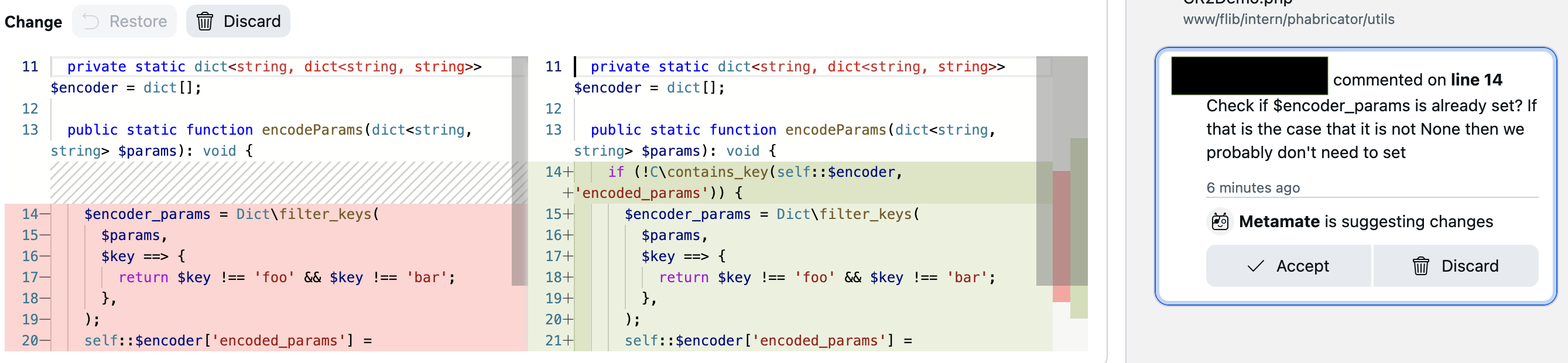}}\hfill

\subfloat[\SftLarge is finetuned on a much larger set of code review comments and the internal codebase. The model not only learned to use Hack functions but also simplified the code by taking advantage of the modern coding patterns offered by the Hack language's null coalescing operator, \ie \texttt{??}. The model learned to adhere to \company's coding guidelines by observing a variety of code review comments and patches that address those comments. ]{\label{largeSFT-response}\includegraphics[width=.95\linewidth]{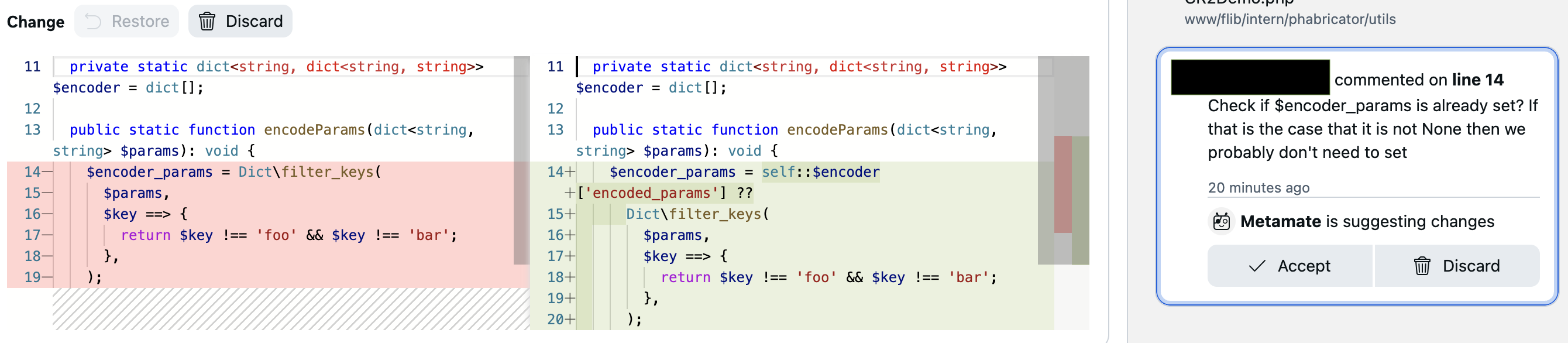}}

\label{figIllustrationModels}
\end{figure*}

\section{Background}
\label{sec:background}
\label{secValidators}

At \Meta, software engineers submit tens of thousands of pull requests (known as ``diffs'') every day. Each diff goes through extensive automated testing, code review, canary deployment, and more to maintain a high level of code quality and reliability in our systems. 
Authoring code in the company is shaped by the scale and complexity of the codebase, which is one of the largest codebases in the industry\footnote{More info at the Meta Tech Podcast: ``What is it like to write code at Meta?'', \url{https://engineering.fb.com/2023/09/05/web/what-like-ship-code-meta-tech-podcast/}. }. Most of the company's code lives in a monorepo, enabling engineers across different teams to collaborate efficiently and make changes fast and with confidence. On the one hand, this setup offers support for shared ownership and visibility --- developers can easily understand how their code interacts with others' code. On the other hand, changes in the code may have quite a big impact on the code of other projects. Hence, before any code is merged, it needs to be vetted by peers, going through rigorous pre-comit code reviews, which are ingrained in the software development culture of the company. 

This deployment strategy has cultivated a routine of frequently ``pushing'' new code to production, supported by a robust continuous integration infrastructure. Prior to any push, the code undergoes peer review, in-house user testing, and comprehensive automated and canary tests. These validators can be broadly categorized into three types: linters, tests, and build checks. Each type plays an important role in maintaining the integrity and consistency of the code base.

\paragraph{Linters} are static analysis tools that examine the source code for potential errors, bugs, or style issues. They help identify and flag problematic code patterns, syntax errors, or deviations from coding standards. Typically, linters are used to:
\begin{itemize}
    \item Enforce coding standards: Ensure that all code adheres to established coding conventions, such as naming conventions, indentation, and commenting.
    \item Detect syntax errors: Identify and report syntax errors, such as missing semicolons, mismatched brackets, or undefined variables.
    \item Flag security vulnerabilities: Detect potential security risks, such as cross-site scripting vulnerabilities.
    \item Improve code readability: Suggest improvements to code organization, structure, and formatting to enhance readability and maintainability.
\end{itemize}
\paragraph{Tests} are an essential component of the validator ecosystem. They verify the correctness and functionality of the generated code by executing it against a set of predefined test cases. Tests can be further divided into:
\begin{itemize}
    \item Unit tests: Focus on individual components or units of code to ensure they function as expected. Unit tests typically involve testing a single function, method, or class in isolation.
    \item Integration tests: Validate how different components interact with each other. Integration tests focus on testing the interactions between multiple units of code, such as APIs, databases, or file systems.
    \item End-to-end tests: Simulate real-world scenarios to verify the overall functionality of the system. End-to-end tests typically involve testing the entire application, from user input to output, to ensure that it works as expected.
\end{itemize}
\paragraph{Build Checks} are a set of automated processes that verify the generated code can be successfully compiled and built. These checks include:
\begin{itemize}
    \item Syntax checking: Verifies that the code adheres to the programming language's syntax rules.
    \item Dependency resolution: Ensures that all required dependencies are correctly resolved and included in the build process.
    \item Code signing: Validates the authenticity and integrity of the generated code.
\end{itemize}
By employing this multi-faceted approach to validation, leads to a codebase, both generated and manual code, that is both reliable and maintainable.

Code review, testing, linting, and build checks all occur in Phabricator.\footnote{Phabricator is an open source project available at \url{https://www.phacility.com/phabricator/}} Code review comments are presented inline and can be seen in Figure~\ref{figIllustrationModels} and in the development of \crres in Section~\ref{secSystem}. A detailed discussion of code review of Phabricator and code review at \Meta can be found in related works~\cite{NudgeMeta,LChen2022FSE-industry}.

\section{Model development} 
\label{sec:modelDevelopment}
\label{LlamaSFT}

In this section, we explain the models we developed for solving the patch generation task from code review comment and the scaling needed for large-scale Supervised Fine Tuning (SFT). We provide details about the process involved in curating fine tuning data sets, training the models, and evaluation results.

\subsection{\gpt} 
Before we developed our own models for solving the patch generation task, we wanted to establish a baseline by testing against a state-of-the-art code generation model. This helps us better understand the success and failure patterns of public models that do not have any internal context about \company. 

As \gpt \cite{openai2024gpt4ocard} was one of the best performing models on various coding benchmarks such as MBPP \cite{austin2021programsynthesislargelanguage}, HumanEval \cite{chen2021evaluatinglargelanguagemodels}, SWE-bench \cite{jimenez2024swebenchlanguagemodelsresolve}, we selected it for our baseline public model.

\subsection{Base Llama model for fine tuning} 
We use a LLaMA 3.1 70B~\cite{grattafiori2024llama3herdmodels} 
as our base model. We align the Llama model for patch generation tasks at \company. Llama comes in four model sizes: 8B, 70B, and 405B parameters. We take the 70B model as our base model because it helps us strike a balance between accuracy, fine tuning cost, and end-to-end inference latency requirements imposed by a scalable \crres system.

\subsection{\SftSmall: Supervised Fine Tuned model with small data set} 
\label{LlamaSmallSFT}

Our patch generation model involves Continual Pre-Training (CPT) on \company's internal code \cite{ke2023continualpretraininglanguagemodels} and Supervised Fine Tuning (SFT) with a small human labeled data set \cite{parthasarathy2024ultimateguidefinetuningllms}. To train the \SftSmall model we initialized it with Llama model weights and continued to pre-train the model on first-party code data. This data includes code used internally such as Hack and Flow. This helps the model learn the basics of coding patterns, frameworks, nomenclature used internally. The details of these internal models have been discussed in related works~\cite{murali2024aiassistedcodeauthoringscale,dunay2024multiline}.

The novel aspect of this work is the Supervised Fine Tuning (SFT) on $\langle review\_comment, patch \rangle$ pairs. These are not synthetically generated data points but mined from \company's code review data. First, we mine a random sample of $\langle review\_comment, patch \rangle$ pairs. We exclude diffs that had failed a lint, test, or build and in the end had 5K data points. In this case, the review\_comment is left by the human reviewer and the patch is the patch pushed by an engineer in response to that code review comment. 

To create a high quality benchmark, we had two domain experts annotate each of the 5K $\langle review\_comment, patch \rangle$ pairs. The annotators were instructed to check the following properties of a data point:

\begin{itemize}

\item \textbf{Actionability.} Does the code review comment require a coding patch fix?

\item \textbf{Instruction quality.} Is the code review comment providing clear instruction about the fix the reviewer is expecting to see happen?

\item \textbf{Accuracy.} Does the published patch fix in the subsequent diff address the code review comment accurately? Since it is common for the developers to write patches that address multiple review comments together, it is also the annotator's job to isolate the code in the patch that addresses the comment. 

\end{itemize}

The domain expert annotators curated a data set with \textit{2.9K high-quality data points}. 
Since the dataset has internal discussion of patch fixes, we cannot release the data publicly. We do not consider the benchmark to be a research contribution, but we do consider the process description to be a contribution. This dataset was used to fine-tune the 70B base Llama model and resulted in the \SftSmall.

\subsection{\SftLarge: Supervised Fine Tuning with a large data set based on human and AI labeled data} \label{LlamaLargeSFT}

To improve the performance of Llama models by better aligning with the task of generating sound patches in response to code review comments, we further scaled the size of the SFT data sets. We employed contract workers with coding expertise as annotators to expand our dataset using the process described in the previous section. Over a 14 week period, we had an average of 7 annotators labeling data. In total, 18k of $\langle review\_comment, patch \rangle$ pairs.

Since human annotation is an expensive and time-consuming task, we develop a binary classifier to further increase our dataset. SFT data sets need to be of high-quality but there is leeway when the dataset size is large enough to compensate for a slight drop in quality~\cite{brown2020languagemodelsfewshotlearners}. We develop a classifier using the following steps.

\begin{enumerate}

\item We picked 7.5K data points randomly from the human labeled data set (18K data points) and fine tuned a Llama 8B model to act as a classifier that can classify a given data point as good (fit to be included in the large SFT data set) or bad (unfit to be included in the large SFT data set). 

\item We measured the precision, recall, and accuracy of this 8B model in classifying the data points correctly by testing it on a data set of 700 data points which are picked again randomly from the 18K human-labeled data set that did not have any overlap with the 7.5k training set. This 8B model records a precision of 0.88, recall of 0.75, and accuracy of 79\%. 

\item Encouraged by the ability of this 8B model to act as a classifier, we scaled the data labeling activity to generate an additional 46K data points. In contrast to the 14 weeks needed to go from 2.5k to 18k datapoints by paid annotators, this model took less than one day to scale to 48k $\langle review\_comment, patch \rangle$ pairs.

\item We used the combined data set of 64K data points, 18K human-labeled data points plus 46K model-classified data points, to perform fine tuning on top of the base 70B Llama model to produce Llama-\SftLarge. We also used a larger context size of 128K compared to 16K of Llama-\SftSmall.

\end{enumerate}

\section{Data, Evaluation Methodology, and Experiments}
\label{secDataEvaluation}

In this section we discuss our benchmarks and offline metrics to determine how well the model might work in production. We then describe our safety trials and experiments that roll \crres into production and ensure that it is effective and does not have any regressions in how long code reviews take to perform. 

\subsection{Offline Evaluation Methodology} \label{OfflineEval}
In the subsequent sections, we generated a dataset to use in supervised fine-tuning of our models. In this section, we discuss a test benchmark to evaluate each of the models. Benchmarking and evaluation are essential tools for guiding model improvements through better training data selection, prompt engineering and supervised fine-tuning. Internal benchmarking is especially vital for understanding the performance of the current models, comparing different models (large versus small, catch-all models versus expert models), and even catching future regressions in production deployments of models.

While there exist many program repair and patch generation benchmarks \cite{jimenez2024swebenchlanguagemodelsresolve, jain2024livecodebenchholisticcontaminationfree, rashid2025swepolybenchmultilanguagebenchmarkrepository, silva2024repairbenchleaderboardfrontiermodels}, they are mostly centered around open source (OSS) projects and repositories. These OSS benchmarks do not necessarily capture the patterns, depth and complexity involved in generating patches and addressing code review comments at a large enterprise such as \company. Therefore, we curated a benchmark that represents the breadth and complexity involved in addressing various code review comments at \company.

Typically, a benchmark comprises two primary components: ground truth data and metrics. Ground truth data can be human-curated or can be generated programmatically. Human-curated data sets tend to have higher quality. 
Generally, for applications such as AI-program repair or patch generation, data set quality is of utmost importance (compared to quantity). Therefore, we pursued the route to generate high-quality but low-quantity benchmarking data sets.

To generate the bespoke human-annotated benchmarking data set, we have senior engineers at \Meta annotate the quality of code review comments and the corresponding patches that were pushed to address those comments. First, we mine a random sample of $\langle review\_comment, patch \rangle$ pairs from \company's code review data. Then, we follow the pruning process explained in Section \ref{LlamaSmallSFT} to prune the space and generate high quality benchmarking data sets.

Using this process, we generated a data set with \textit{206 high-quality data points}. This size of dataset is on par with other AI evaluation benchmarks; for example, Google had 100 manually curated $\langle review\_comment, patch \rangle$ pairs~\cite{google2024cr2}.

To evaluate the \crres system in generating the accurate patches that address code review comments, we ran the system against the benchmark mentioned above. 
The evaluation pipeline provides the \crres system with the original source code file content, a specific code review comment, and the lines of code the comment refers to. The \crres system then generates a patch to address the code review comment. We use the line diff format \cite{muennighoff2024octopackinstructiontuningcode} to generate the patches \ie the patch generated by the \crres system consists of the patch and the line numbers indicating where it should be applied. Once the units of changes with the patch and the line number are generated, the \crres system applies the patches to the original file and regenerates the next version of the file (with the patch applied).

We measure the following metrics to understand the efficacy of various models in generating accurate patches in response to code review comments:

\begin{itemize}
    \item \textbf{Exact Match (EM)} is a simple and strict metric that checks if the file content after applying the AI-generated patch is exactly matching the final version of the file that the humans pushed in response to the code review comment. In theory, a program can be written in many ways. At a simple level, the AI-generated program might differ from the ground truth (human-generated program) in variable names only (while the logic is still the same). However, with EM, we check for strict matches (including program components such as variable names). Therefore, EM is extremely restrictive and can be considered as a lower bound when checking the accuracy of AI-generated patches.
    \item \textbf{Successful Patch Generation (SPG)}: this metric measures if model was able to generate a patch in the first place and if the system was able to apply the AI-generated patch to the program. A patch cannot be generated or applied for various reasons: LLM call errors, line number mismatches in the generated patch, etc. Measuring this metric helps us in understanding the failure scenarios that are not necessarily patch accuracy-related.
\end{itemize}

\subsection{Methodology for Production Experiments}
\label{secMethodOnline}

\begin{table}
\centering
\caption{We conduct four production experiments. The first two experiments are safety experiments and are used to ensure that none of our review time metrics regress and that AI patches are accepted. The second two experiments are rolled out into production for all diffs because the approach is safe and the offline results are strong.}
\begin{tabular}{ p{.25\imagewidth} | p{.18\imagewidth} | p{.18\imagewidth} | p{.18\imagewidth} | p{.18\imagewidth} }

{\bf Experiment}	&	{\bf Expt. 1}	&	{\bf Expt. 2}	&	{\bf Expt. 3}	&	{\bf Expt. 4}	\\ \hline
Experiment type	&	Safety Trial	&	Safety Trial	&	Full Prod Rollout	&	Full Prod Rollout	\\ \hline
Model	&	No AI vs \gpt 	&	No AI vs \gpt	&	Small-LSFT	&	Large-LSFT	\\ \hline
Experiment Date	&	Oct 17 to Nov 21	&	Nov 22 to Dec 20	&	Dec 18 to Jan 14	&	Feb 1 to Mar 2	\\

\end{tabular}
\label{figExperimentSetup}
\end{table}



In this section and in  Figure~\ref{figExperimentSetup}, we discuss the details and timeframe of the four experiments that we run in production with 10's of thousands of engineers. We defer the discussion of the \crres architecture, implementation, and user experience to Section~\ref{secSystem}.

Meta has sophisticated infrastructure that allows engineers and data scientists to run randomized controlled trials~\cite{NudgeMeta,jadad2007randomized}. These trials use the same methodology as those used to run in medical trials of new drugs~\cite{jadad2007randomized}. Our first two experiments are safety trials. In a safety trial the goal is to check that the intervention does not have any negative side effects. In our case, we compare a ``No AI patches control" condition with a ``\gpt AI suggestions" test condition. We monitor how long reviews take to ensure that we do not inadvertently make the review experience worse. As we discuss in results Section~\ref{secSafetyResults}, our safety trial saw regressions and could not be rolled out to a larger population. Our investigations lead to changes in the user experience and our second safety trail saw no regressions. 

After the completion of a medical trial, it is not possible to monitor the entire population. In contrast, at \Meta, we are able to monitor all reviews and ensure that we do not see regressions. As a result of this monitoring infrastructure, we are able to roll-out model improvements, \ie \SftSmall and \SftLarge, as they become available to the entire engineering population and determine if the offline model results hold in production. This ensures that engineers receive the best offline model in production immediately rather than having to go through a controlled trial where only some engineers are exposed to the latest model. Safety metrics are monitored and if regressions are shown, the model is reverted and investigated. In the next section we discuss our goal and safety metrics. 

\subsubsection{Outcome Metrics for Production Experiments} 
\label{secOutcomeMeasures}

\begin{figure}
\centering
\includegraphics[angle=0, scale=\imagescale]{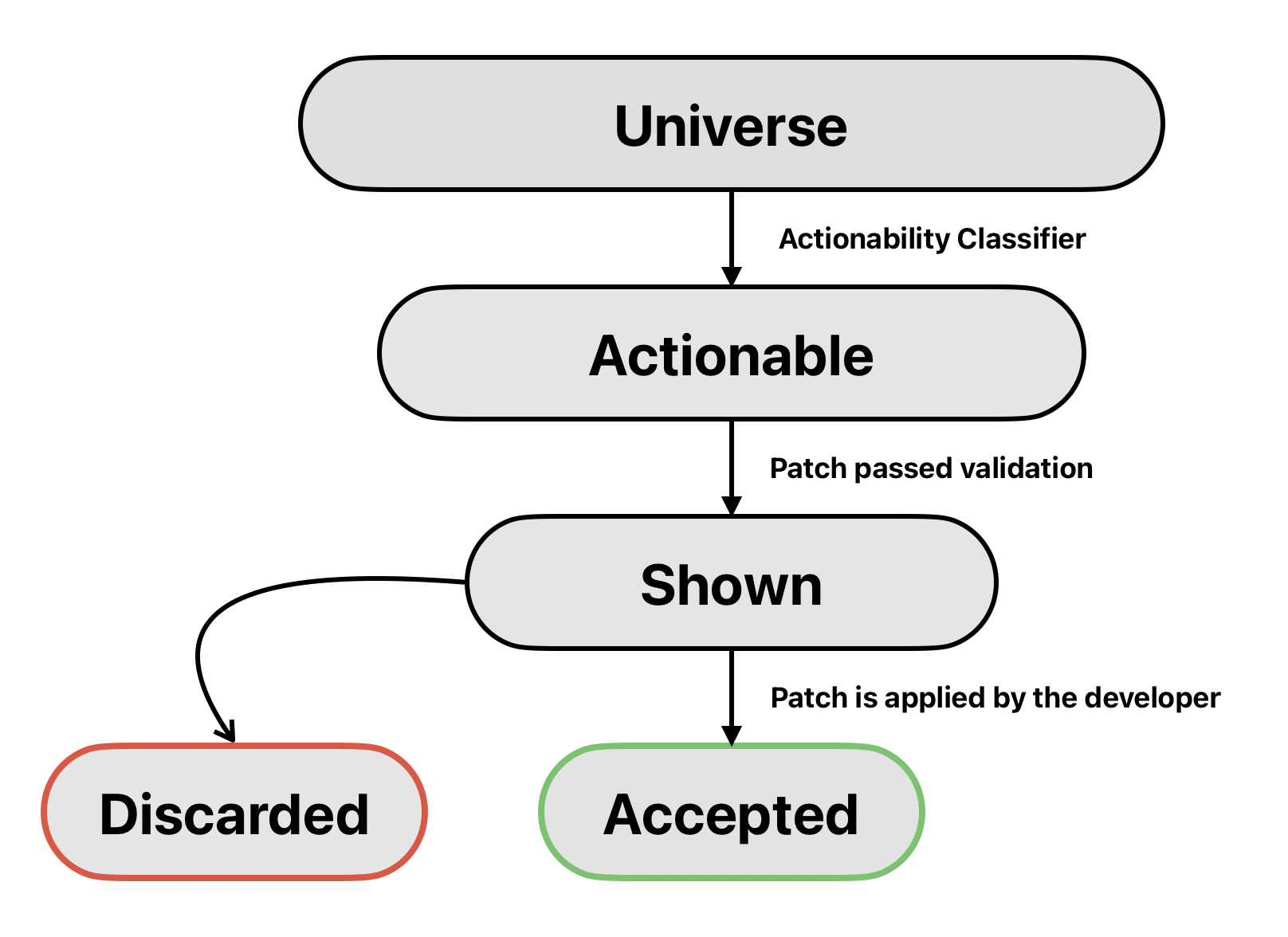}
    \caption{The funnel for \crres. The universe of code review comments at meta is filtered through an actionability classifier. If it is actionable, then validity checks are run, \ie does the AI patch build and pass linters and tests. If yes, then it is shown in the UX and it can be accepted, \ie assisted, otherwise it is discarded.} 
    \label{fig:funnel}
\end{figure}

The data available during production experimentation is shown in Figure \ref{fig:funnel} as the \crres funnel. We define our metrics below.

\textbf{Universe} this is the total set of code review comments the \crres system can address. Note that this set does not necessarily represent all the code review comments at \company. As of writing this paper, 
It includes two of \company's largest code codebases and diffs in four major programming languages.

\textbf{Actionable.} 
We created an In-Context Learning (ICL) \cite{dong2024surveyincontextlearning} solution by instructing a reasonably smart LLM (\gpt in this case) to act as an actionability classifier. We pass a few examples of actionable and non-actionable code review comments as a Few-Shot prompt \cite{song2022comprehensivesurveyfewshotlearning} to perform binary classification. 
We test the resulting ICL-based system on a data set of 240 data points that are manually labeled by human annotators as either actionable or non-actionable. The system recorded a precision of 0.86, recall of 0.85, and F-1 score of 0.85. i.e., the system learned to predict the actionability of a code review comment by causing a very small number of false positives (precision and recall for the actionable class are 0.81 and 0.92 respectively).
The comments that pass the actionability classifier are move to the next stage.

\textbf{Shown.} We do not show all actionable AI-generated patches. To control for quality and accuracy, we show patches that meet one or both of the following: 
\begin{enumerate}
    \item We run validity checks on the AI patch, \ie build, lints, and tests. If it passes all of the checks, then we show the patch. 
    \item Or, if the reviewer explicitly approves the AI patch. 
\end{enumerate}

\textbf{Accepted.} These are the set of AI-generated fixes that are applied by the diff authors:
\begin{enumerate}
    \item The diff author explicitly clicks on the `Accept' button, indicating their approval of the patch, and staging it to be applied on the current version of the source code. Once applied, these approved fixes produce a new version automatically (original code plus AI-generated patch).
    \item We employ several heuristics to detect when the AI-generated patch is applied in the final or committed version of the diff. We do this by first checking for the line numbers of the generated patch in the committed version of the source code file followed by checking for exact match of the code or content at those line numbers. This detection became important when we discovered developers often prefer to copy-paste the suggested patches directly into their IDE. However, this does not fully capture the extent of patch application because developers might move the source code in a file around before or after pasting the AI-generated code in their IDE or rework some parts of the AI-generated code, such as renaming variables.
\end{enumerate}

\textbf{Discarded.} These are the patches that were rejected or otherwise not applied by the diff authors, or which we could not track and attribute the patch application accurately due to the reasons mentioned above.

\subsubsection{Goal and Safety Metrics}

\begin{table*}
\caption{For each experiment, we describe each outcome metric, indicate whether it is a goal metric or a safety metric, and describe the statistical test that is appropriate for the type of data.} 
\begin{center}
\begin{tabular}{ p{.23\imagewidth} | p{.85\imagewidth} | p{.3\imagewidth} | p{.16\imagewidth} }

\hline

{\bf Outcome metrics}	&	{\bf Description}	&	\bf{Experimental Type}	&	{\bf Stat Test}	\\ \hline
Actionable-ToApplied	&	The total number of actionable comments that lead to an AI patch that the author accepts and includes in the codebase.	&	{\bf Goal} in Full Prod Experiment	&	Fisher	\\ \hline
Shown-ToApplied	&	The number of AI patches that a reviewer approves or that pass the validity checks, \ie builds, lints, and tests and are applied by the author.	&	{\bf Goal} in Full Prod Experiment	&	Fisher	\\ \hline
\timeInReview	&	The time when a diff is waiting for review or is being actively reviewed. It excludes the time when the author is reworking the code. We want to make sure that our recommender does not slow down reviewers.	&	{\bf Safety} and Monitored in Full Production	&	t-test	\\ \hline
\eyeballTime	&	The amount of time reviewers spend examining each diff. This is time spent on the review page for a diff (AKA eyeball time)	&	{\bf Safety} and Monitored in Full Production	&	t-test	\\ \hline
\wallClock	&	The wall-clock time from when the diff is published by the author until the review is finished and closed, \ie either accepted or abandoned	&	{\bf Safety} and Monitored in Full Production	&	t-test	\\ \hline

\end{tabular}
\label{tabMetrics}
\end{center}
\end{table*}

Based on the funnel above we define outcome metrics. Table~\ref{tabMetrics} describes our two goal metrics and three safety metrics. 

The total number of actionable comments that
lead to an AI patch that the author accepts and
includes in the codebase is defined as
\begin{equation}
\text{\actionable} = \frac{\text{Applied}}{\text{Actionable}}
\end{equation}
Since this is count data we use a Fisher test to calculate statistical significance.

The number of AI patches that a reviewer approves or that pass the validity checks, \ie builds, lints, and tests and are applied by the author is defined as
\begin{equation}
\text{\shown} = \frac{\text{Applied}}{\text{Shown}}
\end{equation}
Since this is count data we use a Fisher test to calculate statistical significance.

Our safety measures are \timeInReview, which is the amount of time the diff spends in the review state, \ie waiting for review and actively being reviewed. \eyeballTime is the amount of time that the reviewers spend actively looking at the review in Phabricator, \ie eyeball time. \wallClock is the amount of time from when the author publishes the diff until the patch is closed, usually accepted or abandoned, which includes both the \timeInReview and the time the authors spend making the fixes. Each metric is continuous so we use a t-test to calculate statistical significance.

\section{Offline Model Results}
\label{secOfflineResults}

We present the offline results for each of our models.

\subsection{Offline Results for \gpt}
\textit{How well does \gpt generate patches from code review comments at \company?}

The goal of this exercise is to understand how well a state-of-the-art public model, \gpt, performs on patch generation tasks from code review comments using an LLM in a zero-shot fashion. This experiment is our baseline.
We ran an experiment on the benchmarking data set, described in Section \ref{OfflineEval} for the patch generation task. We evaluate against two metrics: exact match (EM) and successful patch generation (SPG) by calling the \gpt in a zero-shot fashion with sample size 1, i.e., we make only one call to the LLM for a given data point or patch generation task. 

As shown in Table \ref{tab:OfflineResults}, the public \gpt model yields an EM score of 59.22\% and a SPG score of 97.57\% in a zero-shot setting. These are impressive results for a public model. \gpt was able to do a great job at generating patches accurately in cases where the source code is written using standard frameworks, following generic coding patterns such as converting a generic \texttt{for loop} into a \texttt{for each loop} or a \texttt{lambda}, fixing minor stylistic changes recommended by the code reviewers, renaming variables, replacing function calls with better versions, etc. Additionally, \gpt was also able to perform reasonably well with code refactoring that involved producing multi-line and multi-hunk changes. 

At the same time, \gpt failed to generate accurate patches for tasks that require deeper knowledge of \company's code base and coding patterns. In Figure~\ref{figIllustrationModels}a, we see that \gpt uses outdated PHP functions rather than the more performant modern Hack functions that are used by \Meta's engineers.

\begin{tcolorbox}
The results for \gpt represent a baseline performance for
a model as it has not been trained on \company's internal data. \gpt generates an exact match 59\% of the time, but is not familiar with the internal coding frameworks, patch formats, code review comment patterns.
\end{tcolorbox}

\begin{table*}
\centering
\caption{The offline results. \SftLarge outperforms \SftSmall and \gpt.}
  \begin{tabular}{lrrrrrr}

\toprule


Model	&	\gpt	&	\SftSmall	&	\SftLarge	\\ \hline
{\bf Offline Evaluation Metric}	&	{\bf Percentage}	&	{\bf Percentage}	&	{\bf Percentage}	\\ \hline
Exact Match (EM)	&	59.22	&	63.11	&	67.96	\\
Successful Patch Generation (SPG)	&   97.57	& 97.09   &   99.51	\\

\bottomrule

  \end{tabular}
  \label{tab:OfflineResults}
\end{table*}

\subsection{Offline Results for \SftSmall}
\textit{How important is fine tuning the model for patch generation from code review comments?} 

To evaluate the impact of continual pre-training (CPT) on internal code corpus and Supervised Fine Tuning (SFT) by generating high-quality human-labeled $\langle review\_comment, patch \rangle$ pairs, we took a public Llama-70B checkpoint and performed CPT and SFT.
Our method is fully described in Section~\ref{LlamaSmallSFT}, to summarize, we examined over five thousand code review comments and the patches that were created to address those comments i.e., $\langle review\_comment, patch \rangle$ pairs. Upon performing a rigorous manual labeling exercise, we were able to generate a data set of 2.9K $\langle review\_comment, patch \rangle$ pairs. We use this dataset to perform SFT.

As shown in Table \ref{tab:OfflineResults}, we see the detailed results of \SftSmall’s performance in patch generation tasks compared to those of GPT-4O, in terms of percentage points (pp) difference. For \SftSmall we see an Exact Match (EM) of 63.11\% and SPG of 97.09\%. 

In Figure~\ref{figIllustrationModels}b, we see that \SftSmall outperforms \gpt when generating patches that involve adding complex type annotations for code written using \company's internal frameworks, predicting accurate functions, performing complex refactoring such as migrating code to use newer and better APIs, modernizing code to follow optimal design patterns, etc.

\begin{tcolorbox}
Supervised fine tuning (SFT) on high-quality human-labeled annotation data sets ($\langle review\_comment, patch \rangle$) sampled from \company's code review data (2.9K data points) allows \SftSmall to produce an exact match of 63\% of the time, an improvement of 3.5pp over \gpt.  
\end{tcolorbox}

\subsection{Offline Results for \SftLarge}
\textit{How important is the scale of the SFT dataset when generating patches from code review comments?}

Encouraged by the success of performing SFT on high-quality task-specific data, i.e., the \SftSmall model, we set out to scale the SFT data generation and further fine tune models on those larger data sets. We call the resulting model \SftLarge, as described in methodology Section \ref{LlamaLargeSFT} which includes 64k $\langle review\_comment, patch \rangle$ training datapoints.

As shown in Table \ref{tab:OfflineResults}, \SftLarge demonstrates substantial gains with respect to its ability to generate accurate patches. \SftLarge records an exact match score of 67.96\%, which is \(\sim \)8 percentage points improvement compared to the public \gpt model. This model also succeeds at the patch generation task (as measured by SPG) in 99.51\% cases, which is \(\sim \)2 percentage points improvement compared to the \gpt model.

The improvements can be attributed to the alignment (SFT) process by curating more diverse examples of $\langle review\_comment, patch \rangle$ pairs and scaling the number of data points used for SFT, i.e. 64K data points versus 2.9K for \SftSmall). In addition to generating the accurate patches, the \SftLarge model learned to align better with respect to generating patches that can be applied to the existing programs (99.51\% for \SftLarge compared to 97.09\% for \SftSmall). Figure~\ref{figIllustrationModels}c shows that \SftLarge not only uses modern Hack functions, but also writes the code in a clean and concise manner that is expected of an engineer.

\begin{tcolorbox}
\SftLarge has an offline exact match of 68\% which is an improvement of 5pp and 9pp over \SftSmall and \gpt, respectively. This result shows the importance of scaling up the training dataset, from 2.9k data points to 64K, using an LLM classifier rather than increasing the dataset with manual human labels.
\end{tcolorbox}

\section{The \crres system} 
\label{secSystem}

We provide an overview of the product experience, various components involved in making the \crres system a reality, and the AI-assisted comment resolution workflow. This helps in developing a deeper understanding of the user experience, workflows, and contextualizing different problems we need to solve to operationalize such a system at scale.

\subsection{System architecture}

\begin{figure}
\centering
\includegraphics[angle=0, scale=\imagescale]{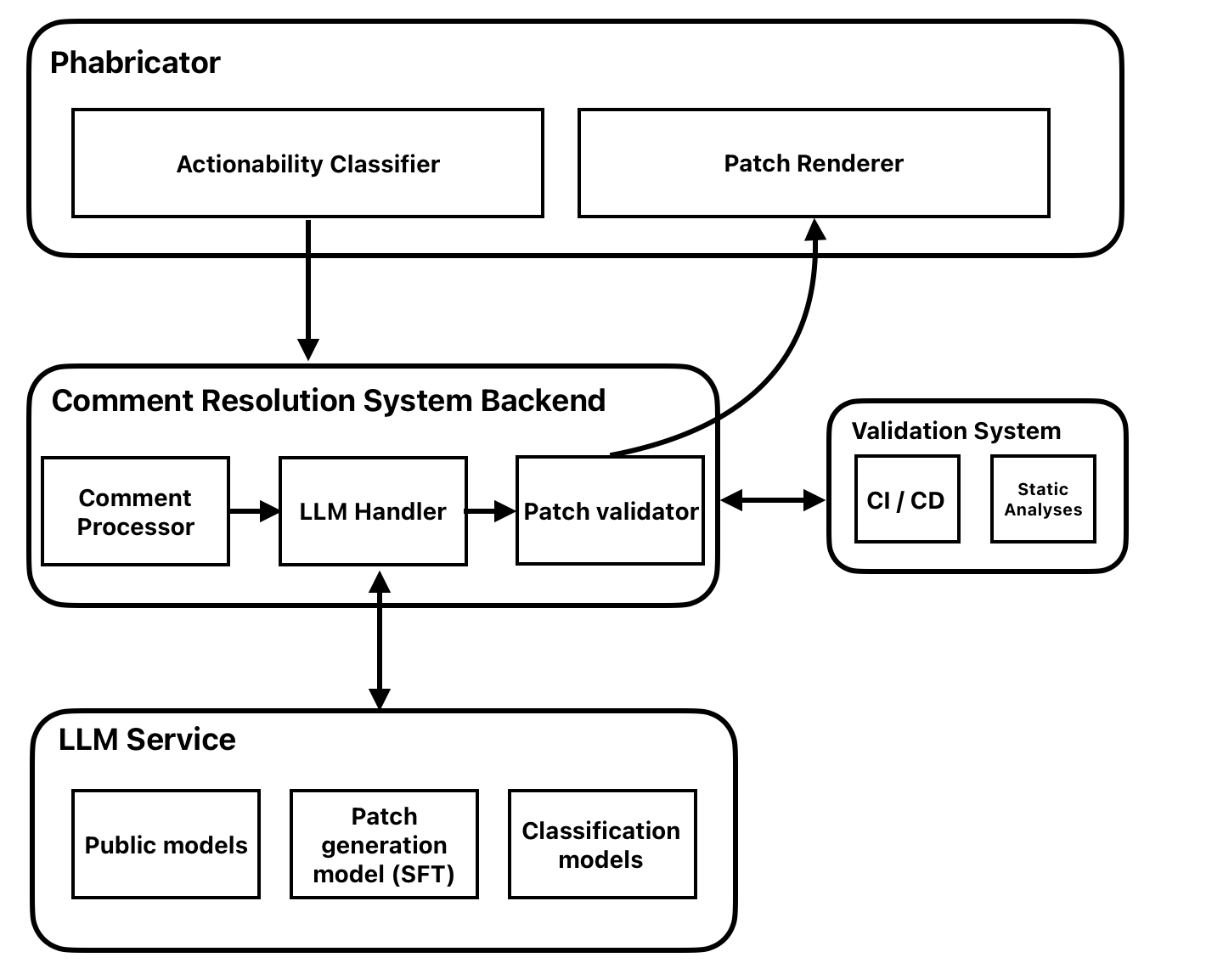}
    \caption{The \crres system architecture}
    \label{fig:architecture}
\end{figure}

The engineering team designed the \crres system with three core components: Phabricator, \crres backend, and our LLM service. As shown in Figure \ref{fig:architecture}, these three independent components work together to generate AI code patches to address code review comments at scale.

\textbf{Phabricator} is the central code review product  which we integrated the \crres system into. We leverage Phabricator for important functionality such as collecting code review comments, executing actionability classifiers, rendering AI-generated patches, and facilitating follow-up actions.

\textbf{\crres backend} is responsible for orchestrating different workflows in the system. These workflows include live, asynchronous processing of code review comments, calling appropriate LLMs for the task, and scheduling validations to check the soundness of the AI-generated patches.

\textbf{Validation system} 
This is an external system that runs a relevant subset of our static analyzers and dynamic verifications, including the linter, tests, and builds described in Section~\ref{secValidators}, that are run on the AI-generated patches. The goal of the validation system is to verify the soundness of the code patches and surface the result as a signal to the diff author and reviewers.

\textbf{LLM Service} hosts different models we fine tuned for patch generation, code review comment classification, that are described in Section~\ref{sec:modelDevelopment} and evaluated offline in Section~\ref{secOfflineResults}. 

\subsection{The User Experience}

\begin{figure*}[ht]
\centering
\includegraphics[angle=0, scale=\imagescale]{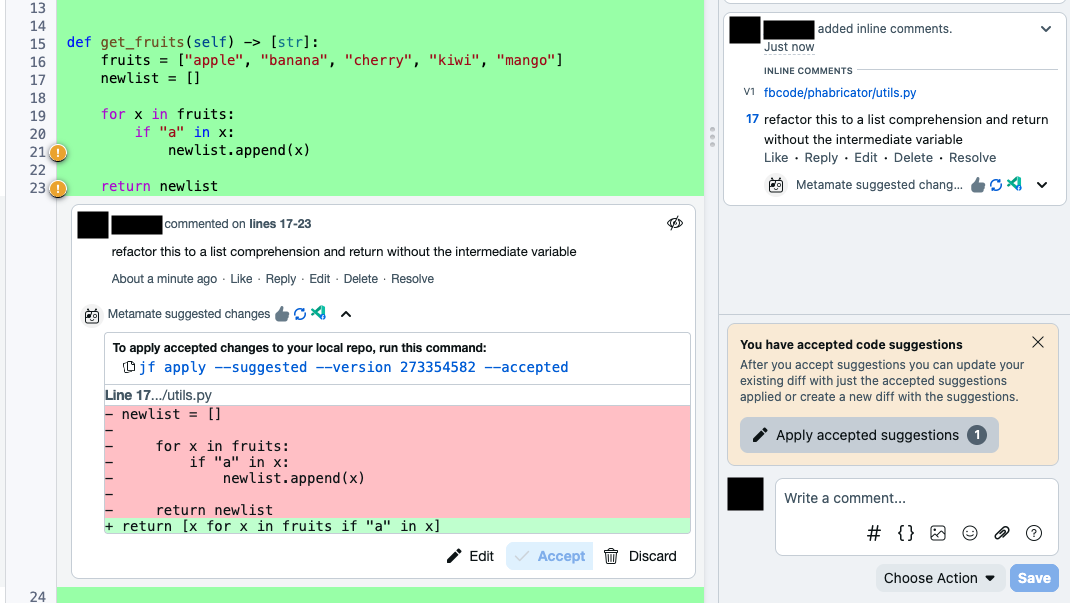}
    \caption{Diff with an accepted code suggestion}
    \label{figProductNormalFlow}
\end{figure*}

Code changes that have been submitted for review in Phabricator are called diffs and are analogous to pull-requests. Engineers leave inline comments on selected ranges of code in a diff. \crres generates a code change suggestion that addresses the comment.
Once the suggestion is generated and it passes the validators, a preview of the code change is shown below the inline comment. Minor edits are visible in their entirety, while more significant changes are abbreviated and can be expanded into a modal dialog for review. Figure~\ref{figProductNormalFlow} shows the author's original code, the reviewers comments, and the AI code suggestions, and the author's acceptance of the AI patch.  

\begin{figure}[ht]
\centering
\includegraphics[angle=0, scale=\imagescale]{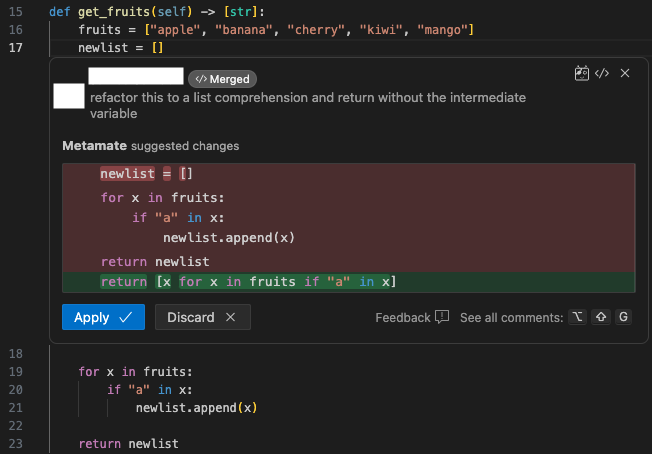}
    \caption{The AI suggestion from the code review comment can be viewed inline in the IDE and allows authors to make additional changes and fixes}
    \label{figProductIdeFlow}
\end{figure}

Accepted suggestions can be applied in several ways, depending on the diff author's preference. 
The code can be accepted in Phabricator and it can be integrated and landed directly without going into an IDE. If the author prefers to view the suggestions in the IDE, they can also be shown inline in the developer's IDE when the associated diff is checked out. This IDE application flow, shown in Figure~\ref{figProductIdeFlow}, is supported through a custom VS Code extension. Available suggestions are displayed to the author when their associated files are opened while working on the diff. The code suggestions can be easily inserted, collapsed, or dismissed entirely.

\begin{figure*}[ht]
\centering
\includegraphics[angle=0, scale=.25]{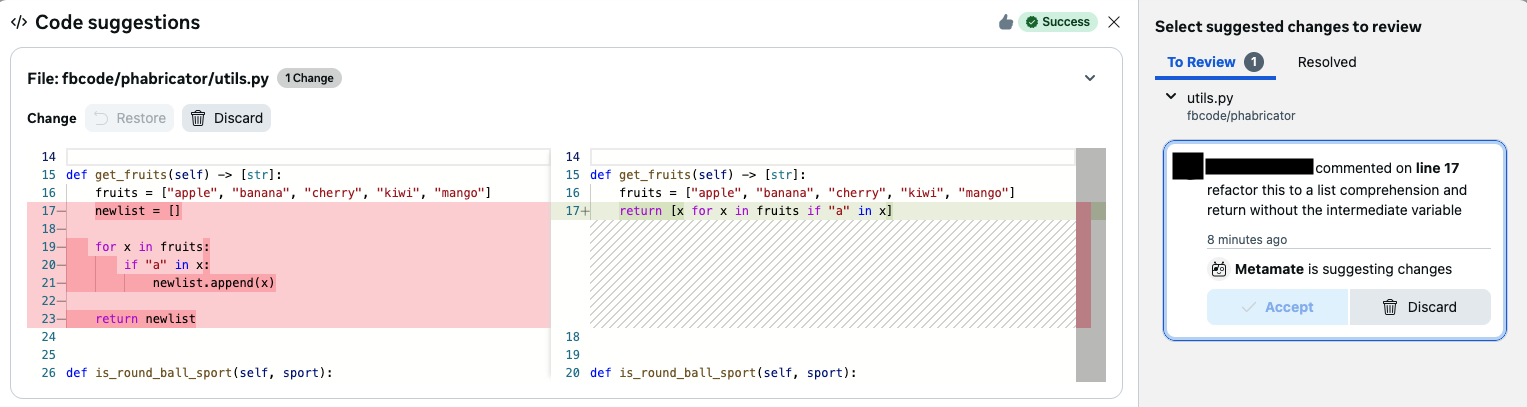}
    \caption{Dialog to review \& apply AI code suggestions}
    \label{figProductModal}
\end{figure*}

Code suggestions can also be reviewed and applied directly to the diff through a modal dialog in Phabricator shown in Figure~\ref{figProductModal}. Authors can click on an inline suggestion to open a dialog containing all active code suggestions linked to the diff, including non-AI suggestions produced by automated linters and human reviewers.
The author can easily navigate between each code suggestion to view the changes in their entirety, and accept the suggestions they want to incorporate, along with any local edits they choose to make. Once the author is satisfied with their selections, they can click a button to immediately publish a new version of the diff with all accepted code changes applied.
Suggestions are automatically archived when they are applied, discarded, or when the code they modify is changed so as to make the suggestion irrelevant. In Phabricator, archived suggestions are marked with the reason why, such as having been merged into a newer version of the diff.

\begin{figure*}[ht]
\centering
\includegraphics[angle=0, scale=.25]{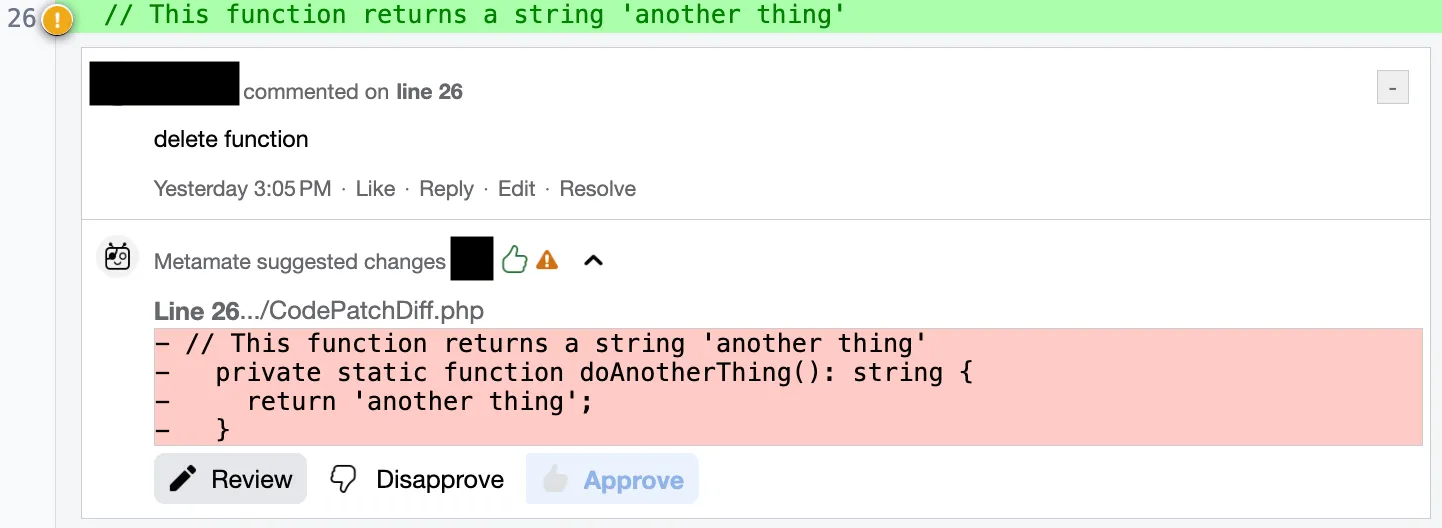}
    \caption{Reviewer approval of an AI suggestion}
    \label{figProductRevApprove}
\end{figure*}

Reviewers may also leave feedback on the AI suggestions as shown in Figure~\ref{figProductRevApprove}. They can choose to endorse the suggestion by approving it, in which case it is featured more prominently to the diff author; or they can disapprove of the suggestion to collapse it for themselves and other engineers.

\section{Safety Trials and Experiments in Production}
\label{secOnlineResults}

In the previous sections of this work, we described our models that in an offline setting successfully generate AI patches from reviewer comments, and we have described the \crres that was built to show the suggestions to reviewers and authors. We now conduct four experiments on engineers at \Meta that write diffs and provide review comments. We described our experimental methodology in Section~\ref{secMethodOnline} and outlined the experiments in Figure~\ref{figExperimentSetup}. The first two experiments are randomized controlled safety experimental trials, while the second two experiments are full production experiments. 

\subsection{Randomized Controlled Safety Trials}
\label{secSafetyResults}

\begin{table}
\centering
\caption{Safety trials. We conduct randomized control trials to determine if the No AI control group vs \gpt AI suggestions lead to any regression in our safety metrics. We see that in Expt. 1, showing the AI patches to reviewers substantially shows down the review process. When the suggestions are collapsed for reviewers in Expt. 2, reviews do not take statistically longer.}

\begin{tabular}{llrrrrrrr}

\toprule

{\bf Experiment}	&	{\bf Expt. 1}			&	{\bf Expt. 2}			\\
Model	&	No AI vs \gpt			&	No AI vs \gpt			\\ \hline
\timeInReview	&	5.5\%	, 	$p = .029$	&	0.42\%	, 	$p = .86$	\\
\eyeballTime	&	6.7\%	, 	$p \ll .001$	&	1.24\%	, 	$p = .66$	\\
\wallClock	&	3.0\%	, 	$p = .095$	&	0.26\%	, 	$p = .67$	\\ \hline

\end{tabular}
\label{tabOnlineSafety}
\end{table}

The goal of \crres is to help authors address reviewer comments more quickly. We want to ensure that AI patch suggestions do not make the review process longer or make the reviewer spend more time during review. Figure~\ref{tabOnlineSafety} shows the results of the safety trials.

\textbf{Expt. 1,} the control condition in the experiment had no AI assisted patches, while the test condition had AI assisted patches. The experiment ran from October 17 to November 21, 2024. We can only calculate the change in safety metrics because the control condition does not have AI patches. 

We see statistically significant regressions in the time a diff was in the review state (5.5\%, $p = .029$) and the time that reviewers looked at the diff (5.5\%, $p = .029$). The total time in review that includes both reviewing and author fixes was not statistically significant but was trending upwards (3.0\%, $p = .095$). This trend likely indicates that while we are slowing reviewers down authors are not impacted. Given these regressions \crres was rolled back and a detailed investigation of the data and qualitative feedback groups was conducted.

The analysis shows longer review times when more AI code suggestions are displayed, even when code suggestions are only viewed without being accepted or rejected. Qualitative feedback from engineers suggests the current experience compels reviewers to read and vet any AI patches prior to accepting a diff, thus increasing review times. 

\begin{tcolorbox}
To summarize, reviewers were not only reviewing authors' code, but also reviewing and feeling a responsibility to check the AI patch that was generated based on their comments. This leads to longer review processing times.
\end{tcolorbox}

\textbf{Expt. 2,} the team collapsed the AI generated patch when the reviewer was viewing the diff. This allowed reviewers to focus on reviewing, but reviewers could still see the generated change if they were interested, see Figure~\ref{figProductRevApprove}. 
We conducted another controlled trial between November 22 and December 18, 2024, with the control being no AI-suggested patches and the test being the new user experience with the author seeing the suggested patches, but the patch being collapsed for the reviewer. 
%
We do not see any statistically significant regressions with the corresponding p values: $p = .86$, $p = 0.67$, and $p = .67$.

\begin{tcolorbox}
To summarize, showing the AI patch only to authors and collapsing it for reviewers, led to a safe user experience that did not have any regressions in the time it takes for code review.
\end{tcolorbox}

\textbf{AI Patch Acceptance for Expt. 1 and 2}

\begin{table*}
\centering
\caption{We compare the \actionable and \shown rates for each experiment and report statistical comparisons between the progressive models.}

\begin{tabular}{lrrrrrrr}

\toprule

{\bf Experiment}	&	{\bf Expt. 1}	&		&	{\bf Expt. 2}	&		&	{\bf Expt. 3}	&		&	{\bf Expt 4.}	\\ 
Model	&	\gpt	&	{\bf ---vs---}	&	\gpt	&	{\bf ---vs---}	&	\SftSmall	&	{\bf ---vs---}	&	\SftLarge	\\ \hline
\actionable	&	10.71\%	&	$p = .75$	&	10.53\%	&	$p \ll .001$	&	16.13\%	&	$p \ll .001$	&	19.75\%	\\
\shown	&	19.90\%	&	$p = .56$	&	21.27\%	&	$p \ll .001$	&	27.55\%	&	$p = .054$	&	28.74\%	\\ 

\end{tabular}
\label{tabOnlineAcceptance}
\end{table*}

In the controlled experimental trials, we established that the new UX and AI generated patches did not degrade or regress the code review process. We now turn to our goal metrics. For both our goal was to assist authors with AI-patches and we quantified how often an actionable suggestion was accepted, \actionable, and how many of those that are shown actually get accepted, \shown. We see an \actionable rate of 10.71\% vs 10.53\%, for Expt. 1 vs Expt. 2 respectively. The corresponding values for \shown is 18.90\% vs 21.27\%. 

We conducted a Fi sher test, neither of these differences metrics show a statistically significant change with  $p = .75$ and $p = .56$. This importantly shows that although we no longer show the changes to reviewers, the authors still accept the AI-patches at a similar rate.

\subsection{Llama-based SFT models in Expt. 3 and 4}

Concurrent to the safety trials, the team responsible for tuning the model conducted offline tests to develop fine tuned models, see Section~\ref{secOfflineResults}. 
%
Unlike medical trials~\cite{jadad2007randomized}, we are able to monitor the entire population of code reviews to ensure and continuously evaluate our goal and safety metrics~\cite{NudgeMeta}. As discussed in our methodology, Section~\ref{secMethodOnline}, we do not conduct a controlled trial because we want all engineers to experience the improvement that we have discovered in the offline tests as soon as possible. We continue to monitor the safety metrics, and can stop any rollout that leads to a safety regression.

\textbf{Expt. 3,} by mid-December the \SftSmall model was trained and showed promising offline results, see Section~\ref{LlamaSmallSFT}. We rolled it out to the entire population of engineer's diffs from December 18, 2024 to January 14, 2025. We find no regressions in our safety metrics including the amount of time it takes to do a review. We see substantial and statistically significant improvements in both the \actionable rate (16.13\%, $p \ll .001$) and \shown rate (27.55\%, $p \ll .001$). This improvement is a 5.6 pp and 6.3 pp increase over \gpt.

\textbf{Expt. 4,} by February the \SftLarge model was trained and showed strong offline results, see Section~\ref{LlamaLargeSFT}. From February 1 to March 2, 2025, we rolled out \SftLarge to all code reviews. We find no regression in our safety metrics. We see a substantial improvement in \actionable rate (19.75\%, $p \ll .001$) a 3.62 pp and 9.22 pp improvement over \SftSmall and \gpt, respectively.
However, we did not see any statistically significant change in \shown (28.74\%, $p = .054$) compared to \SftSmall. 
We suspect that this is because the number of diffs that are actually shown in the UX was similar for both. Given the higher offline accuracy of \SftLarge in future we plan to experiment with showing a larger number of AI suggested patches.

\begin{tcolorbox}
In monitored production rollouts of \SftSmall and \SftLarge, we see substantial improvements over \gpt, 5.6pp and 9.22pp in \actionable, respectively. We monitor our safety metrics and do not experience any regressions.
\end{tcolorbox}

\section{Threats to Validity}
\label{sec:threats}

\subsection{Generalizability}

We have evaluated our Code Review Comment Resolution system only on an internal dataset. This may not be representative of general code reviewing activity in other software development scenarios. Particularly, the specific set of code review tools, internal processes, and company culture might have influenced our experiment. It is unclear if our results could be generalized to other industrial settings beyond \company.

However, the software systems under study cover millions of lines of code and 10s of thousands of developers who are both collocated and working at multiple locations across the world.
We also cover a wide range of domains from user facing social network products and virtual and augmented reality projects to software engineering infrastructure, such as calendar, task, and release engineering tooling. 

\subsection{Construct Validity}

One potential construct validity threat to our study is the evaluation metrics we used to assess the models' performance. We used Exact Match (EM) and Successful Patch Generation (SPG) rates for offline evaluation, and \actionable and \shown for online evaluation. We also captured safety metrics. A potential threat is the possibility of unmeasured confounding variables that may have impacted our results. For example, there may be other factors that influence code review comments and their resolution, such as code quality or team dynamics, that were not captured in our analysis.

\subsection{Internal Validity}
Our study focused only on fixing code review comments that are addressable by suggesting code changes. We did not consider other types of code review comments, such as question-answers, code pointers, knowledge transfer, and general discussions. Therefore, our approach may not fully capture the complexity of the process and may not be suitable for all code reviewing scenarios. Furthermore, our models were trained on historical data, which may not be representative of future code software development scenarios. Therefore, the performance of our models may degrade over time and require periodic retraining.


\section{Literature and Discussion}
\label{sec:literatureAndDiscussion}

Code review and code inspection was first formalized in 1976 by Fagan~\cite{Fagan1976IBM} where the author, reviewers, and mediators met in person to examine completed work artifacts. The inspection was conducted over weeks and sometimes months. 
Much of the early work simply changed the process, often adding more rigidity through process(\eg \cite{knight1993improved}). 
However, Porter \etal~\cite{porter1998understanding} demonstrated that formality of the review process simply extended time in review without identifying additional defects. They showed that {\it expertise} was the most important factor in identifying defects. Votta~\cite{Votta1993Note} proposed asynchronous tool supported review and Perry \etal~\cite{Perry2002TSE} demonstrated that this lighter-weight process found the same number of defects as conducting reviews in meetings. In these late 1990's experiments the review time was on the order of a week. 

Rigby \etal~\cite{rigby2008open,rigby2014peer} showed how effective the extremely lightweight, hugely iterative code review process was on major open source projects. On the Linux and Apache project the review time was on the order of a day. These results generalized to industry with reviews being conducted in less than one day at Microsoft~\cite{rigby2013convergent}. 
At \Meta~\cite{NudgeMeta}, reviews happen very quickly, often being completed in a few hours after publishing the diff (median 2.5 hours), and with around 75\% being completed in under 24 hours.

Recent works on code review have focused on improving efficiency and ensuring there are expert reviewers. For example, nudging overdue reviews improved cycle time~\cite{nudge2023,NudgeMeta}, workload aware recommenders~\cite{whodo2019,Haraji2024TSE} chose the top expert who had the lowest workload, and CommentFinder~\cite{commentfinder2022} ranked the most important comments to focus on during review. 


With the rise of generative models (LLMs), the domain has shifted towards automating more complex activities. Cihan et al.~\cite{cihan2024automated} propose a method to do LLM-based code review and evaluate it in an enterprise setting. AutoTransform~\cite{autotransform2024} proposed a neural-network based method for transforming code during the code review process.

More closely related to our paper is the line of work on suggesting fixes for code review comments left by reviewers. In this area, CodeReviewer~\cite{zhiyu2022automating} pre-trains a multi-task model on several code reviewing tasks, including diff quality estimation, review generation, and code review comment resolution. At \company, code review cannot be skipped and engineers still must review and comment on diffs, so we cannot use the first two approaches in this paper. Although the datasets and models are different, our offline results indicate an exact match of 67\% compared to the top results of 30\% reported by Zhiyu \etal. Furthermore, we present results for the deployment of the system to engineers and the feedback they provide. 

A closely related work is Google's code review comment resolution system~\cite{google2024cr2} that is also deployed at scale. Our paper is similar in principle, but differs in the important ways: (i) we compare the performance of different models on the problem, including both external closed and open-source models, (ii) we show the impact of human (and subsequently, LLM) annotation on the models' performance, and (iii) rather than a confidence measure, we utilize an actionability classifier that decides whether to invoke the system on a review comment. 

A direct comparison is difficult due to the different methodologies, models, and outcome measures, however, we attempt some reasonable comparisons. In an online setting, our \SftLarge model had an \actionable rate of 19\%, while Google reported the number of changes applied by the author divided by the number of changes that Google's model was confident in at 15\%. Another comparison is the number of Google's changes that the reviewer approved relative to those applied by the author at 22\% compared to our number of changes \shown rate of 28\%. We are {\it not} claiming to outperform the Google model, but instead to be performing at a similarly high level.

Broader related work includes the recent body of work on fine-tuning LLMs for various software engineering tasks across the SDLC, ranging from code authoring to risk assessment~\cite{murali2024aiassistedcodeauthoringscale,dunay2024multiline,maddila2024aiassistedsqlauthoringindustry,abreu2024drs}. These works have shown that fine-tuning LLMs for specific tasks is quite effective, especially when working in an organization's code base. For certain tasks, prompting a general purpose model can indeed go a long way towards solving the problem. However, in this paper, we show that it is not the most effective solution, by comparing prompting a general model (\gpt) and fine-tuning specialized models. Of course, there arises a need to balance between model effectiveness versus proliferation of too many specialized models across an organization. The problem of ``generating code patches from natural language'' might offer a good balance here, but it is left as future work to explore.

Recent studies have shown an improvement in productivity of developers at using GitHub in terms of the number of commits as well as in a controlled experiment of 96 engineers at Google. However, we are unaware of studies that have examined the impact on code reviewers. 

Peng \etal~\cite{peng2023impact} showed an 56\% increase in task completion time using co-pilot on 95 developers. Cui \etal~\cite{cui2024effects} found that Copilot increased the number of diffs written by a developer by 25\% compared to the control group that did not have access to Copilot. Paradis \etal~\cite{paradis2024dgoogle} present an experiment involving 96 engineers that showed a statistically significant decrease in time on a specific task from 114 without AI to 96 minutes with an AI code assistant. However, when other factors such as tenure and amount of time usually spent programming, the impact of AI was no longer statistically significant. We are unaware of work, especially at the industrial scale, that examine the productivity gains from generating fixes from comments. The code comment fixer at Google~\cite{google2024cr2} explicitly asks reviewers to check the generated fix code, but does not quantify the impact on the time that an author must spend doing the review. Indeed our findings demonstrate that showing these fixes to reviewers actually slows them down, and so they are collapsed and only shown to authors. This finding echoes the difficulties with AI-code assistants that when the generated code is large or incorrect it actually makes more work for code authors~\cite{dunay2024multiline,murillo2023if} to code reviewers.
%
\section{Concluding Remarks}
\label{sec:conclusion}
%
%
\crres has enjoyed a wide and consistent adoption among employees at \company. As of writing this paper, \crres is enabled for engineers at \Meta across four major programming languages. The \crres system processes thousands of code review comments every week and generates patches for the ones deemed actionable by the actionability classifier.

The \crres system generates a patch for a code review comment that is classified by our actionability classifier as `actionable'. Then, a patch is shown to the diff authors only if the AI-generated patch passes our post-processing checks and is deemed to be accurate and useful by our addressability filtering.
As shown in Table \ref{tabOnlineAcceptance}, the \crres system powered by public \gpt model recorded an accepted rate of 10.5\% on all the actionable code review comments and 21.2\% accepted rate on all the patches that were shown to the diff authors. The same system when powered by \SftSmall and \SftLarge models recorded accepted rates of 16.1\% and 19.7\%, respectively, on all the actionable comments. Similarly, \SftSmall and \SftLarge recorded accepted rates of 27.5\% and 28.7\%, respectively, on all the comments shown to the diff authors respectively.

We conducted safety trials to refine our user experience and ensure that we did not see regressions in how long reviews take. In our safety trial we saw statistically significant regressions on how long reviewers looked at a diff and how long it took to close a diff. Reviewers were taking on the additional role of checking the fix to the code rather than commenting and making suggestions to the author. 
At Google, the integration of AI-assisted code fixes has shown great promise; but the burden is likely significant~\cite{google2024cr2}. As engineers write comments, AI-generated fixes are immediately shown to reviewers, who then approve them before they are shared with authors. While Google reported on the frequency of reviewer approvals and author acceptance, there is still room for exploration --- specifically, how this process affects review time. A key question emerges: ``should reviewers take on the responsibility of fixing authors' code?'' Our findings suggest that perhaps not. Not only could this increase review time, but it also adds to the already significant workload of experienced reviewers~\cite{Haraji2024TSE}. Collapsing the AI patch for reviewers solved the regression in time in review, and appropriately, put the burden of fixing the code back on authors. 
The work around code review and AI is moving at a fast rate, we are excited to see other advances by researchers and in industry. 

\section*{Acknowledgements}
We would like to thank Cindy Ding, Larry Lai, Daniel Cheng, Victor Montalvao, Matthew Bessey, Monisha Shivakumar, Imad Ahmad, Michael Jiang, Shahin Sefati, Patrick Riggs, Anita Anderson, Kristian Kristensen, Ale Contenti, and Killian Murphy for their help and support with this work.

\bibliographystyle{IEEEtran}
\bibliography{main}

\end{document}